\documentclass[AMA,LATO1COL]{WileyNJD-v2}
\usepackage{moreverb}

\newcommand\BibTeX{{\rmfamily B\kern-.05em \textsc{i\kern-.025em b}\kern-.08em
T\kern-.1667em\lower.7ex\hbox{E}\kern-.125emX}}

\articletype{Article Type}%

\received{<day> <Month>, <year>}
\revised{<day> <Month>, <year>}
\accepted{<day> <Month>, <year>}

\usepackage{hyperref} 
\usepackage{amssymb} 
\usepackage{xspace}
\usepackage{color}
\usepackage{xcolor}
\usepackage{fancybox}
\usepackage{fancyhdr}
\usepackage{ifthen} 
\usepackage{paralist} 

\usepackage{amsmath,amsfonts}
\usepackage{float}
\usepackage{graphicx}
\usepackage{textcomp}

\usepackage[algo2e,ruled,vlined,linesnumbered]{algorithm2e}

\definecolor{mygreen}{rgb}{0,0.6,0}
\definecolor{mygray}{rgb}{0.5,0.5,0.5}
\definecolor{mymauve}{rgb}{0.58,0,0.82}

\usepackage{listings} 

\usepackage[T1]{fontenc}
\usepackage{courier}
\usepackage{multirow}

\usepackage{flushend} 

\newboolean{acmtemplate}
\setboolean{acmtemplate}{false} 
\newboolean{anonymous}
\setboolean{anonymous}{false}
\newboolean{includecomments}
\setboolean{includecomments}{true}

{{\ttfamily \hyphenchar\the\font=`\-}

\newcommand{\paperTitle}{Redefine Paper Title Later}
\newcommand{\submitTo}{ACRONYM of conference later}
\newcommand{\maxPagesAllowed}{?? to be redefined}

\clubpenalty = 10000
\widowpenalty = 10000
\displaywidowpenalty = 10000

\hypersetup{
    colorlinks,%
    citecolor=black,%
    filecolor=black,%
    linkcolor=black,%
    urlcolor=gray,
    linktocpage=true,
    bookmarks=true,
    bookmarksopen=true,
    pdfpagemode=UseOutlines,
    pdftitle={\paperTitle},    
    pdfauthor={Anonymous},     
    pdfsubject={Draft for \submitTo Conference},   
    pdfcreator={PDF LaTex},   
}

\setlength{\textfloatsep}{2mm} 

\ifthenelse{\boolean{anonymous}} {

}{

}

\ifthenelse{\boolean{includecomments}} {
    \newcommand{\nb}[3]{
    	{\colorbox{#2}{\bfseries\sffamily\scriptsize\textcolor{white}{#1}}}
    	{\textcolor{#2}{\sf$\blacktriangleright$\textit{#3}$\blacktriangleleft$}}}
} {
    \newcommand{\nb}[3]{}
}

\newcommand{\Mehrdad}[1]{\nb{Mehrdad}{red}{#1}}

\newcommand{\Someone}[1]{\nb{\Someone}{olive}{#1}}
\newcommand{\Ebert}[1]{\nb{Ebert}{cyan}{#1}}

\newcommand{\sa}{\textsc{Small-Amp}\xspace}
\newcommand{\py}{\textsc{Pynguin}\xspace}

\newcommand*{\RQOne} [1] {Isn't this a splendid research question here ?}
\newcommand*{\RQTwo} [1] {And isn't this one even better ?}
\newcommand*{\RQThree} [1] {This one tops it all, doesn't it ?}

\begin{document}
\renewcommand{\paperTitle}{AmPyfier: Test Amplification in Python}
\renewcommand{\maxPagesAllowed}{20}
\title{\paperTitle}

\author[1]{Ebert Schoofs*}

\author[1,2]{Mehrdad Abdi}

\author[1,2]{Serge Demeyer}

\address[1]{\orgdiv{Dept. Computer Science}, \orgname{University of Antwerp}, \orgaddress{\state{Antwerp}, \country{Belgium}}}

\address[2]{\orgdiv{AnSyMo}, \orgname{Flanders Make vzw}, \orgaddress{\state{Antwerp}, \country{Belgium}}}

\corres{*Ebert Schoofs, Dept. Computer Science, University of Antwerp, Antwerp, Belgium. \email{ebert.schoofs@student.uantwerpen.be}}


\abstract[Abstract]{
Test Amplification is a method to extend handwritten tests into a more rigorous test suite covering corner cases in the system under test.
Unfortunately, the current state-of-the-art for test amplification heavily relies on program analysis techniques which benefit a lot from explicit type declarations present in statically typed languages like Java and C++.
In dynamically typed languages, such type declarations are not available and as a consequence test amplification has yet to find its way to programming languages like Python, Ruby and Javascript. 
In this paper, we present AmPyfier, a proof-of-concept tool, which brings test amplification to the dynamically typed, interpreted language Python.
We evaluated this approach on 7 open-source projects, and found that AmPyfier could successfully strengthen 7 out of 10 test classes (70\%).
As such we demonstrate that test amplification is feasible for one of the most popular programming languages in use today.
}


\jnlcitation{\cname{%
\author{Schoofs E.}, 
\author{M. Abdi }, and 
\author{S. Demeyer}} (\cyear{2021}), 
\ctitle{\paperTitle}, \cjournal{Journal of Software: Evolution and Process. Special Issue: Automatic Software Testing from the Trenches}, \cvol{TODO}.}

\maketitle


\section{Introduction}
As proven over and over again, faults are often more expensive to fix than to prevent.
In order to prevent regression and detect bugs, an adequate testing suite is necessary.
To determine the program-based adequacy of a test suite, multiple criteria exist such as, but not limited to, code coverage (method/brach/statement)~\cite{zhu1997software} or the mutation score~\cite{Papadakis2019}.
Manually writing a completely adequate test suite is a labour and time-consuming task.
This is where research concerning test amplification and generation proves to be useful~\cite{lu2020semi, fraser2011evosuite}.

Test amplification and generation tools respectively try to automatically improve a test suite, or generate one based on given criteria.
Their goals are similar, but the difference between amplification and generation is in the main input it takes.
While test generation focuses only on the program under test, test amplification generates tests based on modification or extension of the existing handwritten testsuite~\cite{danglot2019snowballing}.

In order to automatically improve the mutation coverage of unit tests, Baudry et al. created DSpot, a test amplification tool for the statically typed language Java~\cite{baudry2015dspot}.
Danglot et al. demonstrated the effectiveness of DSpot, by running it on 10 mature open source projects.
In their experiments, DSpot successfully improved 26 out of the 40 ($\approx$65\%) test classes under study~\cite{danglot2019automatic}.
DSpot is able to leverage the fact that Java is a statical typed language, and typing information can be deduced from a static analysis of the source and test code.

For dynamically typed languages, test amplification techniques cannot solely rely on static analysis.
To deduce the type of variables and values that are to be asserted or, in order to support type-sensitive input amplifiers, dynamic techniques are needed.
Abdi et al. have made a proof of concept tool, \sa with which they showed that test amplification is also possible for the dynamically typed language Pharo Smalltalk~\cite{abdi2019test, abdi2021smallamp}.
They introduced the concept of dynamic type profiling, with which they were able to overcome the lack of type information in the source code.
While it is certainly interesting from an academic point of view that test amplification can be extended from statically, to dynamically typed languages, no other tools have been developed or presented to the best of our knowledge.

As dynamically typed languages, such as Python, become increasingly popular, adequate testing suites are needed.
Indeed, the popularity of Python is still on the rise\cite{srinath2017python}, owning to its ease of learning, huge amount of external libraries, and the further rise of data-mining and AI research.
This leads to more and more projects being written in Python, or parts of it reliant on code written in Python.

In 2020, Lukasczyk et al. introduced a tool, capable of test generation for Python, \py \cite{lukasczyk2020automated}.
\py, however, leverages the techniques used in statically typed languages through the assumption that the system under test contains type information with Python's type annotations.
Auger~\cite{auger} is another tool that automatically generates tests for Python, but only considers the given execution of the project under test.

In contrast to the popularity of Python, the research into automatically improving a test suite for a Python project is still sparse, and especially in regard to test amplification, non-existent.

With this paper, we extend the concept of test amplification to Python, the most popular (dynamically typed) language today according to IEEE\footnote{\url{https://spectrum.ieee.org/top-programming-languages/}}, and present \textbf{AmPyfier}\footnote{\url{https://ansymore.uantwerpen.be/artefacts/ampyfier}}, a proof of concept tool.
Python is a dynamically typed interpreted language, with a less strict syntax as for example Smalltalk.
Furthermore, despite Python being an Object-Oriented programming language, it supports multiple programming paradigms.
Python allows a developer to disregard the object-oriented design and write code following the procedural and functional paradigms.\cite{van2007python}

We furthermore demonstrate that it is not only feasible, but also effective for existing open-source projects.
In our evaluation on 7 open-source projects including 10 test classes, AmPyfier successfully amplified 7 test classes (70\%), with one case showing an increase up to 53.06\% in the mutation score.
The experimental data is publicly available in our GitHub repository\footnote{\url{https://github.com/SchoofsEbert/AmPyfier_evaluation}} where the generated reports, and the amplified test suites can be found.

In the following section, we provide some background, what exactly Test Amplification is and how it is implemented in DSpot, and \sa.
Furthermore, we give some more background about Python, unit-testing in Python, and related work in regard to Test Generation for Python.
In Section~\ref{sec:AmPyfier}, we explain the main algorithms of AmPyfier itself with the help of a running example, and how the implementation for Python differs from \sa and DSpot.
We have evaluated AmPyfier on 7 open-source projects, the results are discussed in~\ref{Results}, as well as the lessons we have learned from these experiments.
In the last section, we conclude and discuss further work that needs to be done.

\section{Background and related work}
\subsection{Test Amplification}\label{subsec:test-amplification}

Test amplification can be seen as family of test generation, aimed at improving a Test Suite against a given criterion.
The main difference is, in test generation, the starting point is the project under test, whereas in test amplification, the already existing test suite is taken as input.
By adopting methods like test-driven development (TDD), developers write test code during or before the writing of production code.
Consequently, the majority of modern software projects include a considerable amount of test code.
These test suites often contain meaningful information about the project under test, and what should be tested.
Although these test suites may cover most of the main scenarios in the code, some of the corner cases may still be missed and remain untested.

Test amplification is defined as followed by Danglot et al.~\cite{danglot2019snowballing}:
\begin{quote}
``\textbf{Test amplification} consists of exploiting the knowledge of a large number of test cases, in which developers embed meaningful input data and expected properties in the form of oracles, in order to enhance these manually written tests with respect to an engineering goal.''
\end{quote}

\noindent
They also divide the existing test amplification tools into four categories:
\begin{compactenum}
    \item Amplification by adding new tests as variants of existing ones
    \item Amplification by synthesizing new tests with respect to changes
    \item Amplification by modifying test execution
    \item Amplification by modifying existing test code
\end{compactenum}

\subsection{Test Amplification Components}\label{subsec:test-amplification-steps}
A test amplification tool typically consists of four main components.
In the next parts, we provide short descriptions for each component, and lastly, take a look at related work.

\paragraph{Assertion Amplification}
This module is responsible for adding missed assertions.
A recent assertion amplification technique proposed by Xie~\cite{xie2006augmenting} works based on a dynamic analysis that runs the test to be amplified and captures the object states during the execution.
In this technique the value of getter-methods of an object, or the return values of function calls are captured then extra assertion statements are added.

\paragraph{Input Amplification}
Test input is all statements in a test code except the assertion statements.
With input amplification, new versions of test input are generated based on the existing test input.
The amplified test input may take a different execution path or bring the objects under test into new states.
In other words, the input amplification component empowers a test amplification tool to explore the search space of all possible tests.


DSpot uses an input amplification technique inspired by ETOC~\cite{Tonella_2004}.
In this technique, a test input is modified statically by a set of input amplification operators, and new versions of the test input are generated.
Input amplification operators range from changing the literal values by simple arithmetic operations on integers and modifications of strings to addition, duplication, or removal of method and function calls.
Multiple such amplifiers are applied after each other to combine the different ways in which a test is modified.

\paragraph{Selection}
The goal of test amplification is to improve the tests in regard to a quantifiable engineering goal, expressed as a metric.
Possible metrics are: coverage or mutation score improvement, fault detection capability improvement, oracle improvement, and debugging effectiveness improvement~\cite{danglot2019snowballing}.
Test Amplification tools generate lots of new tests derived from the existing test suite.
The selected metric is calculated for each newly generated test.
If it is able to improve the metric value, the generated test method is selected as an amplified test method otherwise it is discarded.


\paragraph{Type Profiling}
Performing static analysis in dynamically typed languages like Python yields less information than in statically typed languages because the source code in dynamic languages does not include type information.
Adding a new method call is an example of a type-sensitive input amplifier.
To generate new arguments for new method and function calls, typically we need to know which input type to pass to the call.
For statical typed languages, such as Java, we can derive the types of the input through a static analysis of the project under test, but for languages such as Python, we need to derive the types dynamically.
Since the existing test suite is one of the main inputs in test amplification tools, performing dynamic type profiling by running the existing test suite is possible.
The concept, introduced by \sa, is to run the existing test suite and capture (1) the type of arguments in the methods under test (2) the type of variables in the original test methods.

\subsubsection{Related work: DSpot and \sa}
AmPyfier is the adoption of the algorithms introduced in DSpot and \sa into the Python ecosystem.
Whereas DSpot is developed for the statically typed language Java, \sa performs test amplification for the dynamically typed language Pharo Smalltalk.
Both amplify the test suite through the addition of new tests as variants of the existing ones.
New tests are generated based on the existing tests and added to the test suite based on a given acceptance criterion.

The main amplification algorithms used in DSpot and \sa, and now AmPyfier are quite similar.
They repeat the input amplification, assertion amplification, and selection steps iteratively for all existing test methods.
Mutation Coverage is proposed as a selection criterion in DSpot, and is taken over by \sa.
AmPyfier uses a multi-level metric for selection whereas it detects the improvements based on code coverage and mutation testing.
In other words, it computes the mutants only in the covered parts of code, and a generated test method is selected if it is able to increase either test coverage or killed mutants.

To know the values to be asserted in the assertion amplification step, DSpot and \sa both use the same technique.
First, the statements of interest are encapsulated in observation statements, and the test is executed.
The results are collected and, based on the observations, the assertion statements are constructed.
AmPyfier tackles this problem in a similar manner, albeit with a different implementation technique.
It leverages the fact that Python is interpreted and uses python \texttt{sys.settrace()} function.
An in depth discussion can be found in subsection~\ref{subsec:assertion-amplification}.

For gathering the type information, \sa introduces a dynamic type profiling step before the main loop of the algorithm.
To extract this information \sa adds Metalinks on all variables in a test method.
Metalink\cite{costiou2020sub} is a fine-grained reflection mechanism, that allows to install AST node level proxies.
Subsequently, \sa runs the test.
The meta-links are fired when the variables are used to capture the type-information in the run-time.
This type information is then used, during Input Amplification, to select another variable in the test to use as input for the newly added method.\cite{abdi2021smallamp}
AmPyfier however, is able to leverage the Python Interpreter instead of Metalinks to gather such information.
See~\ref{subsubsec:type-profiler} for a more in depth explanation.

\subsection{Python}\label{subsec:python}
Python is a dynamically typed interpreted language, which supports multiple programming paradigms.
Python doesn't force developers to write their code in an object-oriented manner and encapsulate all functionality in classes, allowing developers to write code following multiple programming paradigms.
In the background, however, everything in Python is an object, even the current stack frame.
This fact is leveraged by the Python Tracer, which allows to dynamically obtain all necessary information about what is happening in the project under test during execution.

While everything is an object according to the inner workings of Python, Python has no notion of encapsulation.
Private or protected attributes can not be enforced.
Every attribute is public, and can be accessed from everywhere.
It is up to the developer to follow the coding conventions where an attribute should be considered protected if it is prefixed with one underscore (e.g. \texttt{\_protected\_attribute}), and private if it is prefixed with two underscores (e.g. \texttt{\_\_private\_attribute}).

Not only attribute access is very relaxed, but Python also has a very relaxed import system.
Different import statements can be used for the same import behaviour.
It is possible to only import a single variable, function, or class from a module, or to import a whole package consisting of multiple files.
Such a package needs to have an \texttt{\_\_init\_\_.py} file, which can contain code, or import statements on its own.
Furthermore, it is not necessary to define all the imports at the beginning of the file.
It is possible to import modules from inside functions or even loops for example.
The code in an imported module or package is always executed upon import.
This flexible import system is challenging for test amplification because we must establish the class under test for a given test case.

It is possible in Python to pass a custom traceback function to the Python interpreter with the \texttt{sys.settrace()} function.
Those custom trace functions make it possible to investigate the execution of a Python project, and collect runtime information before certain events occur, such as a new line, a function call, or the occurrence of an exception.
The current stack frame, as well as the event type and a possible argument, are passed to the custom trace function and can be inspected.
From this stackframe, for example, it is possible to access the variables currently in scope.
The \texttt{inspect} module elevates this further and provides several useful functions to help get information about live objects such as modules, classes, methods, functions, tracebacks, frame objects, and code objects~\cite{inspect}.
For example, it allows to get all members of an object with the \texttt{getMembers()} function, or the name of a module some function was defined in using \texttt{getModule()}.
The \texttt{inspect} module can even return the specific source code of an object or method definition with \texttt{getsource()}.

Python is an idiomatic language~\cite{knupp2013writing}, for doing a specific task, there is only one typical, well-known and optimal way.
Other ways are not preferred even if they are grammatically correct.
Since following the idioms makes the code more readable and easier to maintain, it is important for tools like AmPyfier, which synthesize code portions, to support such guidelines.

\subsubsection{Unit testing in Python}
Unit testing in Python has no straightforward conventions.
Since Python projects don't have to follow object-oriented design, a unit can be seen as both a module or a class.
Additionally, Python has multiple unit test frameworks, which can function completely differently.
Two of the most popular are the default \texttt{unittest} framework included in the Python Standard Library, and Pytest\footnote{\url{https://pytest.org}}.

Unittest is inspired by the unit testing framework for Java, JUnit.
All test methods should be contained in a class derived from the \texttt{unittest.TestCase}.
A single test file can consist of multiple such classes.
A test suite can be considered as a single test file, or a collection of test files.
In this paper, we refer to a Test Suite as one test file, possibly containing multiple test classes.

\subsubsection{Test generation in Python and related work}
In regard to test amplification, no tools have been developed for Python to the best of our knowledge.
On the other hand for test generation, there has already been done some work.

In 2020 Lukasczyk et al. introduced \py~\cite{lukasczyk2020automated}.
A tool that can automatically generate unit tests for Python.
However, \py implements the test generation techniques of whole-suit generation\cite{fraser2011evosuite} and feedback-directed random generation\cite{pacheco2007feedback}, established for statically typed languages and uses them on Python.
Therefore, \py is most effective under the assumption that the system under test contains type information with Python’s type annotations.
It thus does not completely address Python's dynamic nature. 

Another tool that automatically generates unit tests for Python is Auger~\cite{auger}, introduced in 2016.
Auger leverages the Python Tracer and tracks function calls related to the module under test.
If a function is called that is defined in the module under test, Auger keeps track of both the values of the arguments as well as the return values.
Based on those values, assertions can be generated.
Auger thus does not need to rely on type annotations, but only generates tests for the given execution of the project under test.
If a function is not called, it won't be tested.
Furthermore, it doesn't handle exceptions very well.

While both \py and Auger could be used as a good starting point to develop a test suite, they both have their limitations.
\py relies on type annotations, whilst Auger does not test non-executed code.

\section{AmPyfier}\label{sec:AmPyfier}
In this section, we present AmPyfier.
In section~\ref{subsec:running-example} we present the running example, used to explain the working of AmPyfier in the subsequent subsections.
A broad overview of the main algorithm is given in section~\ref{subsec:logic}.
Section~\ref{subsec:discovering-the-module-under-test} explains how AmPyfier discovers the module under test despite Python's flexible import system.
In sections~\ref{subsec:assertion-amplification} and~\ref{subsec:ia}, the main differences in assertion and input amplification for Python are set out.
Lastly, the selection of tests is described in section~\ref{subsec:selection-of-amplified-methods}

\subsection{Running Example}\label{subsec:running-example}
To explain the workings of AmPyfier, we use a running example,  namely a simple fund implementation: SmallFund.
See listing~\ref{lst:class} for the source code of SmallFund.
The SmallFund class in our running example has five public methods; \texttt{get\_balance}, \texttt{deposit}, \texttt{is\_empty}, \texttt{get\_transactions} and \texttt{get\_self}.
Furthermore, it has two protected attributes (\texttt{\_balance} and \texttt{\_transactions}) and one public attribute (\texttt{owner}).

\begin{lstlisting}[language=python,label={lst:class}, caption={The SmallFund class}, float, floatplacement=H!]
class SmallFund:
    def __init__(self, owner):
        self._balance = 0
        self._transactions = []
        self.owner = owner

    def get_balance(self):
        return self._balance

    def deposit(self, amount):
        if amount >= 0:
            self._balance += amount
            self._transactions.append( amount)
        else:
            raise Exception("Can not deposit negative amounts")

    def is_empty(self):
        return self._balance == 0

    def get_transactions(self):
        return self._transactions

    def get_self(self):
        return self
\end{lstlisting}

The manually created test suite for SmallFund is presented in listing~\ref{lst:suite}.
This test suite covers the main scenario of depositing positive amounts.
However, it does not cover corner cases like depositing an $amount$ less than zero.

\begin{lstlisting}[language=python,label={lst:suite}, caption={Original test suite}, float, floatplacement=H!]
import unittest
from SmallFund import SmallFund

class SmallFundTest(unittest.TestCase):
    def setUp(self):
        self.b = SmallFund("Iwena Kroka")

    def testDeposit(self):
        self.b.deposit(10)
        self.assertEqual( self.b.get_balance(), 10)
        self.assertIsInstance( self.b.get_self(), SmallFund)
        self.b.deposit(100)
        self.b.deposit(100)
        self.assertEqual( self.b.get_balance(), 210)
\end{lstlisting}

\subsection{Logic}\label{subsec:logic}
The logic of AmPyfier is shown in Algorithm~\ref{main_logic}.
The main algorithm is inspired by the work of DSpot, and also makes use of type profiling similar to \sa.
The default input is a single Python test file, the Test Suite (TS), the list of amplifiers/mutators (A), and the number of subsequent amplifier runs ($n$).

\begin{algorithm2e}
    \caption{AmPyfier logic}
    \label{main_logic}
    \SetKwInOut{Input}{input}
    \SetKwInOut{Output}{output}
    \SetKwFunction{FindModuleUnderTest}{FindModuleUnderTest}
    \SetKwFunction{AssertionAmplify}{AssertionAmplify}
    \SetKwFunction{InputAmplify}{InputAmplify}
    \SetKwFunction{TypeProfile}{TypeProfile}
    \SetKwFunction{Score}{Selection}

    \Input{Test Suite TS}
    \Input{List of amplifiers A}
    \Input{Number of Input Amplifier runs $n$}
    \Output{Amplified Test Suite ATS}

    ATS $\leftarrow \emptyset$\;
    \For{TC in TS}{
        $mut$ $\leftarrow$ \FindModuleUnderTest{TC}\;
        ATC $\leftarrow$ TC\;
        $score$ $\leftarrow \Score($ATC, $mut)$\;
        \For{$test$ in TC}{\label{line:innerloop_begin}
            $a\_test$ $\leftarrow$ \AssertionAmplify{$test$, $mut$}\;
            $a\_score$ $\leftarrow \Score($ATC $\cup$ $\{a\_test\}, mut)$\;
            \If{$a\_score$ > $score$}{
                ATC $\leftarrow$ ATC $\cup$ $\{a\_test\}$\;
                $score \leftarrow a\_score$\;
            }
            $tp$ $\leftarrow$ \TypeProfile{TC, test}\;
            IT  $\leftarrow$ \InputAmplify{$a\_test$, A, $n, tp$}\;
            $score$, IIT $\leftarrow \Score($ATC $\cup$ IT, $mut)$\;\label{score}
            ATC $\leftarrow$ ATC $\cup$ IIT\;
        }\label{line:innerloop_end}
        ATS $\leftarrow$ ATS $\cup$ ATC\;
    }
\end{algorithm2e}

AmPyfier iterates over each test class (TC) in the Test Suite and finds the module under test ($mut$) for this Test Case\footnote{Since all test classes are inherited from the class  \texttt{unittest.TestCase}, we use the name test class and test case interchangeably.}.
This module can be a single Python file, or a package consisting of multiple files.
Afterwards, AmPyfier scores the current Test Case against the module under test using the multi-level coverage calculator.
The variable \texttt{score} contains a tuple of absolute coverages.
We explain it in more details in section~\ref{subsec:selection-of-amplified-methods}.

The inner loop (lines~\ref{line:innerloop_begin} to~\ref{line:innerloop_end}) loops over each test in the Test Case, and Assertion Amplifies it.
If the assertion amplified test ($a\_test$) improves the current score, it is added to the Amplified Test Case (ATC).

After a test case is assertion amplified, the original test is passed to the Type Profiler in order to dynamically obtain information about the various types used in the method or function calls in this test.
The amplified test, the amplifiers to use, and the number of input amplification iterations and also the profiled type information are passed to the input amplifier.
The input amplifier, which encompasses an assertion amplification run at the end, returns a set of input amplified test methods (IT).

Finally, the input amplified tests are scored, and the current score is updated (line~\ref{score}).
The input amplified tests that improved the current score are appended to the amplified test case.
Once each test method is amplified, the amplified test case is added to the amplified test suite.
At the end, each amplified test case is added to the Amplified Test Suite, which is returned after each Test Class is amplified.

AmPyfier currently only supports the \texttt{unittest} framework.
The reason why \texttt{unittest} is preferred to be supported first is: (1) it is the standard testing library proposed by the language,
(2) it complies with xUnit testing practices similar to JUnit in Java and sUnit in Smalltalk.
However, the tool is developed in an extensive way that support for other frameworks, such as \texttt{Pytest}, can be added.

\subsection{Discovering the module under test}\label{subsec:discovering-the-module-under-test}

The first task AmPyfier needs to solve is to determine the module under test for each individual test class.
Because of Python's very flexible import system, as discussed in~\ref{subsec:python}, this is no obvious task.
AmPyfier tackles this problem using the following technique:

\begin{compactitem}
\item
First, an inspection of the imports of the test class is performed and imported modules belonging to the project under test are stored.
\item
Subsequently, all function calls and constructors are observed.
The module that is used the most in the test case is considered the module under test.
\end{compactitem}

In case AmPyfier fails to detect the right module under test, AmPyfier reports on this, and the whole class is skipped.
The module under test can also be provided as a configuration argument.
In the case the module under test is passed as an argument, it is used for the whole test suite.

\subsection{Assertion Amplification}\label{subsec:assertion-amplification}
Algorithm~\ref{alg:aa} shows how Assertion Amplification is implemented in AmPyfier.
It consists of four steps: 
(1) firstly, the assertions are stripped from the test.
(2) secondly, the test without assertions is executed and observations are constructed.
(3) then, execution and collecting the observations are repeated F times.
These observations then are compared and all non-deterministic observations are discarded.
(4) Finally, (new) assertions are constructed based on those observations.

\begin{algorithm2e}
    \caption{Assertion Amplification}
    \label{alg:aa}

    \SetKwInOut{Input}{input}
    \SetKwInOut{Output}{output}
    \SetKwFunction{RemoveAssertions}{RemoveAssertions}
    \SetKwFunction{Observe}{Observe}
    \SetKwFunction{RemoveNonDeterministic}{RemoveNonDeterministic}
    \SetKwFunction{AddAssertions}{AddAssertions}

    \Input{Test Method \texttt{test}}
    \Input{Unit under Test \texttt{unit}}
    \Output{Assertion Amplified Test Method \texttt{amplified}}

    amplified $\leftarrow$ \RemoveAssertions{test}\;
    observations $\leftarrow$ \Observe{amplified, unit}\;
    \For{$F$ times }{
    	observations2 $\leftarrow$ \Observe{amplified, unit}\;
	\If{observations $\neq$ observations2}{
        		observations $\leftarrow$ \RemoveNonDeterministic{observations, observations2}\;
    	}
    }
    amplified $\leftarrow$ \AddAssertions{amplified, observations}\;
\end{algorithm2e}

The first step is to remove the original assertions.
The test source is scanned statically and all method calls that are known as an asserting statement are replaced with their asserted expression.

The second step is observing the test method.
The test method is dynamically executed and the values of the variables and the state of the objects are captured.
For observing the values, DSpot and \sa manipulate the source code to inject observation statements.
However, Python is an interpreted language and lots of information is accessible during runtime.
AmPyfier uses Python \texttt{sys.settrace()} function along with a set of custom trace functions to hook in different points during the execution and observe the objects' states and collect the return values of function calls.

The observations of the state of objects are also made possible thanks to the interpreted nature of Python.
Python has a built-in module \texttt{inspect} enabling the possibility to derive all information about live objects during runtime, such as their state, the module and file it was defined in, the name, \textellipsis.
AmPyfier uses a custom method based on \texttt{inspect.getmembers()} to get all the public attributes and methods of an object, and stores them in the observation.
This custom method is exactly the same as \texttt{inspect.getmembers()}, except it adds the possibility to also capture members or variables that cause exceptions instead of simply raising them.
This proves its usefulness after input amplification, where AmPyfier may have pushed an object in an erroneous state.


When AmPyfier extracts an \texttt{assertRaises} statement, the extracted statement will raise an exception.
Exceptions also occur when the test input is mutated by the Input Amplification module.
So, it is crucial for AmPyfier to be able to handle the exceptions correctly during observation of the test method.
When the observation process is interrupted by an exception, AmPyfier finds the line causing the exception.
Then, it wraps this line in a \texttt{Try}-\texttt{Catch} block and restarts the observation process.
This allows AmPyfier to assert raised exceptions on a per-line basis, and still assert the other statements in a test.

To prevent generating flaky tests, AmPyfier excludes all non-deterministic values from the observations.
AmPyfier uses a configurable constant \texttt{F} and repeats the observing process \texttt{F} times.
After each iteration, the observation sets are compared, and the non-deterministic observations are removed.


After the test has been observed and the values to be asserted collected, the final step is to construct the assertion statements.
AmPyfier loops on a line per line basis over the method-body, and constructs the assertions based on the observations originating from that line.

Listing~\ref{lst:asamp} shows our running example test method \texttt{testDeposit} after assertion amplification.

\begin{lstlisting}[language=python,label={lst:asamp}, caption={Assertion Amplified test method}, float, floatplacement=H!]
def testDeposit_amp(self):
    self.b.deposit(10)
    self.assertEqual( self.b.get_transactions(), [10])
    self.assertFalse( self.b.is_empty())
    self.assertEqual( self.b.owner, 'Iwena Kroka')
    self.assertEqual( self.b.get_balance(), 10)
    self.assertIsInstance( self.b.get_self(), SmallFund)
    self.b.deposit(100)
    self.assertEqual( self.b.get_balance(), 110)
    self.assertEqual( self.b.get_transactions(), [10, 100])
    self.assertFalse( self.b.is_empty())
    self.assertEqual( self.b.owner, 'Iwena Kroka')
    self.b.deposit(100)
    self.assertEqual( self.b.get_transactions(), [10, 100, 100])
    self.assertFalse( self.b.is_empty())
    self.assertEqual( self.b.owner, 'Iwena Kroka')
    self.assertEqual( self.b.get_balance(), 210)
\end{lstlisting}

\subsection{Type Profiler}\label{subsubsec:type-profiler}

Before the assertion amplified test is passed to input amplification, it is passed to the type profiler.
Since input amplification is mainly performed statically, some type-sensitive input amplifiers, such as the addition of a new method call, need type information in advance.

Python is a dynamically typed language and does not force developers to add type information to the source code.
Python supports type annotations since Python 3, but there is no guarantee that developers annotate types during development.
In addition, annotations defined for functions and variables are not enforced by the python runtime and they are only used by third-party tools like IDEs and linters, so there is no guarantee that they are defined correctly~\cite{pythonTyping}.
As a consequence, AmPyfier does not rely on the annotated types in the source code and uses a dynamic type profiling step to collect type information.

Dynamic type profiling is introduced by \sa for test amplification in dynamically typed languages~\cite{abdi2021smallamp}.
In simple words, the type profiler exploits the fact that it is possible to extract type information dynamically by executing the existing test suite.
As is mentioned in section \ref{subsec:test-amplification}, the existing test suite is the main input in the algorithm and lots of useful information about the project under test are embedded inside.

The test suite as well as the test method to be amplified are executed in order to detect the type of arguments in the methods and functions under test.
This type profiler works similar to the \texttt{Observing} step in assertion amplification, and leverages the ability to pass custom tracing functions to the Python interpreter through the  \texttt{sys.settrace()} function.
Allowing AmPyfier to monitor calls executed in the test method, and discover the possible types used as input for different method and function calls.
With this information, new method, or function calls can be generated as well as their random inputs.

\subsection{Input Amplification}\label{subsec:ia}

The algorithm for input amplification in AmPyfier is presented in listing~\ref{alg:ia}.
Input Amplification starts similar to Assertion Amplification, namely, stripping the tests of its assertions again.
Then we loop over all the amplifiers $n$ times in order to combine different amplifiers.
This loop generates loads of new tests, so each time we have looped over the amplifiers, only $T$ randomly selected tests are kept.
Those $T$ tests are added to the results, and also passed on to the next iteration.
Once the $n$ iterations are completed, all the selected tests are assertion amplified.
Finally, the assertion amplified tests are sorted on their modification count, and returned.

\begin{algorithm2e}
    \caption{Input Amplification}
    \label{alg:ia}
    \SetKwInOut{Input}{input}
    \SetKwInOut{Output}{output}
    \SetKwFunction{RemoveAssertions}{RemoveAssertions}
    \SetKwFunction{AssertionAmplify}{AssertionAmplify}
    \SetKwFunction{Size}{Size}
    \SetKwFunction{Sort}{Sort}
    \SetKwFunction{SelectRandom}{SelectRandom}

    \Input{Test method \texttt{test}}
    \Input{Number of iterations \texttt{n}}
    \Input{List of amplifiers \texttt{AMPS}}
    \Input{Typing information \texttt{tp}}
    \Output{Input Amplified Tests \texttt{AT}}

    $temp$ $\leftarrow \{~ \RemoveAssertions{test}\}$ \;
    $results \leftarrow \emptyset$\;
    \For{$n$ times}{
        $amplified$ $\leftarrow \emptyset$ \;
        \For{$amp$ in $AMPS$}{
            $amplified$ $\leftarrow$ $amplified \cup amp.apply(temp, tp)$\;
        }
        \If{\Size{$amplified$} > $T$}{
            $amplified$  $\leftarrow$ \SelectRandom{$amplified$, $T$}
        }
        $results \leftarrow results \cup amplified$\;
        $temp \leftarrow amplified$\;
    }
    AT $\leftarrow$ \AssertionAmplify{$results$}\;
    AT $\leftarrow$ \Sort{AT}\;
\end{algorithm2e}

In order to input amplify the test, we need to strip the test from any assertions.
Statements that were asserted in the original test method are extracted, whereas amplified assertions are completely removed.
If the assertions were kept in the test, the test would fail on the amplified inputs.
Furthermore, we want to construct new assertions based on those amplified inputs.

Once stripped from assertions, AmPyfier loops over the list of selected amplifiers $n$ times.
Each time the loop over the amplifiers is completed, new tests are generated, increasing the number of tests exponentially for each loop.
To counter this, after each loop, only a configurable amount of tests ($T$) is randomly selected and passed on to the next loop.
The default value of $T$ is 200.

The amplifiers used in input amplification are based on the ones used in DSpot and \sa:
\begin{compactitem}
    \item Literal Mutation:
    \begin{compactitem}
        \item Numerical Values: $0, +1, -1, *2, /2$
        \item Strings: add random char on empty string, double the string, random substring of half the size, replace with empty string
        \item Booleans: negation
        \item Unification of literals
    \end{compactitem}
    \item Mutation of Method Calls:
    \begin{compactitem}
        \item Removal
        \item Duplication
        \item Addition of a new method call
    \end{compactitem}
\end{compactitem}


In listing~\ref{lst:iasrem}, the assertion amplifed method in listing~\ref{lst:asamp} is shown during stages of Input Amplification.
The listing on the left, is the method after stripping the assertions, and the listing on the right is the method after three transformations:
1) One method call has been removed at line 2, 
2) at line 3, a new call to a random method is added with a random input value of the correct type, 
3) and another call is removed at line 7.
Note that, the unit under test used in this example has only three methods, of which two are getters.
So, the only interesting method which can thus be added, \texttt{deposit()} is generated.

\begin{multicols}{2}[\captionof{lstlisting}{An example of transformations applied during input amplification}]

\begin{lstlisting}[language=python]
def testDeposit_stripped(self):
    self.b.deposit(10)

    self.b.get_balance()
    self.b.get_self()
    self.b.deposit(100)
    self.b.deposit(100)    
    self.b.get_balance()
\end{lstlisting}

\begin{lstlisting}[language=python,label={lst:iasrem}]
def testDeposit_i_amplified(self):
    # removed statement
    self.b.deposit(-45485) # new statement
    self.b.get_balance()
    self.b.get_self()
    self.b.deposit(100)
    # removed statement    
    self.b.get_balance()
\end{lstlisting}
\end{multicols}

After the Input Amplifier loop is finished, all newly generated tests are Assertion Amplified, and sorted depending on the number of modifications:
\[
N_{modifications} = N_{all\_assertions} + N_{transformations} - N_{original\_assertions}
\]

Adding a new assertion statement counts as one modification, the same for any modification made through Input Amplification.
Regenerating the original assertions is not counted as a modification.
The Input and Assertion Amplified test (listing~\ref{lst:iasamp}), in our example, has a modification count of 7: it has 3 transformations, 4 new assertions are added and 3 assertions are regenerated.

Sorting helps AmPyfier to prefer methods that differ the least from the original test method, or those which have the least new assertions.
So, if two different amplified tests kill the same mutant, the test with the lowest modification count is selected.

\begin{lstlisting}[language=python,label={lst:iasamp}, caption={Assertions \& Input Amplified test method}, float, floatplacement=H!]
def testDeposit_amplified(self):
    with self.assertRaises(Exception):
        self.b.deposit(-45485)
    self.assertEqual( self.b.get_balance(), 0)
    self.assertIsInstance( self.b.get_self(), SmallFund)
    self.b.deposit( 100)
    self.assertEqual( self.b.get_transactions(), [100])
    self.assertFalse( self.b.is_empty())
    self.assertEqual( self.b.owner, 'Iwena Kroka')
    self.assertEqual( self.b.get_balance(), 100)
\end{lstlisting}

\subsection{Multi-Metric Selection}\label{subsec:selection-of-amplified-methods}

Input Amplification yields large amounts of new test methods, and most of them do not increase the coverage, thus a method to decide on which tests to add to the Amplified Test Class is needed.
This method scores the current amplified test case with the addition of a newly generated test based on a given adequacy criterion.
AmPyfier mainly uses the same criterion as suggested by DSpot: the number of mutants a test suite is capable to kill.

One of the main downsides of mutation testing is that, naively, a test has to be executed the same number of times as there are mutants.
For test amplification, this means that a test has to be executed the number of mutants times the number of amplified tests.
It comes as no surprise that reducing the number of mutants against which have to be tested can drastically decrease the runtime.

Adopting the experience of mutation testing in Google~\cite{googleMutationSEIP2018}, the tool generates mutants only in the covered lines of code.
However, this practice has as a result that amplified tests that reach previously non-covered code won't be selected, because they do not kill any mutant.
Therefore, AmPyfier introduces multi-metrics selection: the usage of code coverage along with mutation testing.
A test is selected if it increases the code coverage or it kills new mutants.
Furthermore, if an amplified test reaches previously uncovered code, new mutants are generated for those lines.

Python has many capable mutation testing frameworks, but the one used by AmPyfier is \texttt{mutaTest}\footnote{\url{https://github.com/EvanKepner/mutatest}}.
This framework is chosen thanks to the extensive API that allows for the introduction of caching.
AmPyfier thus does not need to retest for every mutant that is already killed and can throw away mutants that result in time-outs, drastically decreasing the runtime of AmPyfier.
To derive the coverage score of a test, AmPyfier makes use of the Coverage.py framework\footnote{\url{https://github.com/nedbat/coveragepy}}.

The return value \texttt{score} is a tuple of the absolute number of covered elements.
The first element of this tuple shows the number of all covered lines, and the second element shows the number of killed mutants.
This variable is used in the main algorithm (Algorithm~\ref{main_logic}) to detect any improvements in the coverage.

\section{Evaluation}\label{Results}

To evaluate AmPyfier, we have run it using the default configuration on multiple test files from multiple small open-source projects found on GitHub.
In total, we have evaluated AmPyfier on 10 test classes as can be seen in Table~\ref{tab:projects}.
Those projects are selected based on recent activity and the fact that they have more than 50 stars on GitHub.
Furthermore, their test suites had to have been developed using Python's standard \texttt{unittest} framework.
The original coverage and mutation scores of the test classes range from a very low score to a very high score.
The goal is to test to which extent AmPyfier is still able to increase the mutation score, even for test suites with an already high mutation score.
All classes without any alive mutants are rejected because they have no room to improve.
The results of our evaluation are publicly available in our GitHub repository\footnote{\url{https://github.com/SchoofsEbert/AmPyfier_evaluation}}.

\begin{table*}[!ht]
    \centering
    \begin{tabular}{l|l|l}
    	
        \textbf{Project} & \textbf{Repository}& \textbf{Selected test class} \\
        \hline
        Python Twelve Tone & \url{https://github.com/accraze/python-twelve-tone} & \texttt{tests.test\_composer.TestMatrix} \\
        \hline
        Pippin Nano Wallet & \url{https://github.com/appditto/pippin_nano_wallet} & \texttt{tests.validator\_tests.TestValidators}\\
         & & \texttt{tests.crypt\_tests.TestAESCrypt}\\
        \hline
        Whois & \url{https://github.com/richardpenman/whois}& \texttt{tests.test\_main.TestExtractDomain}\\
        \hline
        PyPDF2 & \url{https://github.com/mstamy2/PyPDF2}& \texttt{tests.tests.PdfReaderTestCases} \\
         & & \texttt{tests.tests.AddJsTestCase} \\
        \hline
        Addict & \url{https://github.com/mewwts/addict} & \texttt{test\_addict.DictTests} \\
         & & \texttt{test\_addict.ChildDictTests} \\
        \hline
        MPyQ & \url{https://github.com/eagleflo/mpyq}& \texttt{test.test\_mpqarchive.TestMPQArchive} \\
        \hline
        PJ & \url{https://github.com/eatonphil/pj} & \texttt{tests.test\_pj.TestStringMethods} \\\hline
      \end{tabular}
    \caption{Projects amplified with AmPyfier}
    \label{tab:projects}
\end{table*}

The default configuration of AmPyfier is a) the usage of the cache to minimize runtime, b) automatic discovery of the module under test c) 2 observation runs to counter flaky tests d) 200 tests can be collected during each loop inside the Input Amplification part, e) there are 3 of these loops, and f) all amplifiers discussed in section~\ref{subsec:ia} are used.

The results of this experiment are presented in Table~\ref{tab:results}.
The second and third columns (\texttt{\%MSO}, and \texttt{\%MSA}) in this table represent the mutation score before, and after test amplification.
The forth column (\texttt{\%MSI}) shows the increase in the mutation score which is the difference of third and second columns.
The column \texttt{\%RMSI} represents the relative increase in the killed mutants.
\[
\%RMSI = 100 * \frac{\#Mutants.killed_{Amplified} - \#Mutants.killed_{Original}}{\#Mutants.killed_{Original}}
\]
The columns \texttt{\#MO} and \texttt{\#MA} represent the number of test methods in the original test class, and the number of amplified test methods produced by AmPyfier.

\begin{table}[t]
    \centering
    \begin{tabular}{l|c|c|c|c|c|c|c}
      \textbf{Test classes}  & \textbf{\%MSO} & \textbf{\%MSA} & \textbf{\%MSI} & \textbf{\%RMSI} & \textbf{\#MO} & \textbf{\#MA}\\\hline
      TestMatrix & 89.29 & 89.29 & 0 & 0& 5 & 0\\
      TestValidators & 86.57 & 89.55 & 2.99 & 3.45& 2 & 2\\
      TestAESCrypt & 91.43 & 98.57 & 7.14 & 7.81& 1 & 4\\
      TestExtractDomain & 2.69 & 55.84 & 53.15 & 1975.86& 7 & 4\\
      PdfReaderTestCases & 49.73 & 54.08 & 4.35 & 8.75& 2 & 2\\
      AddJsTestCase & 48.08 & 48.08 & 0 & 0& 2 & 0\\
      DictTests & 84.06 & 86.96 & 2.90 & 3.45& 63 & 2\\
      ChildDictTests & 84.06 & 86.96 & 2.90 & 3.45& 63 & 2\\
      TestMPQArchive & 59.47 & 59.47 & 0 & 0& 5 & 0\\
      TestStringMethods & 95.59 & 96.32 & 0.74 & 0.77& 10 & 1\\\hline
    \end{tabular}
    \caption[Caption for LOF]{The result of evaluation AmPyfier on 10 test classes}
    \label{tab:results}
\end{table}


AmPyfier is able to increase the mutation score for 7 out of the 10 classes (70\%).
The increase in mutation score ranges from 0.74\% in a well tested project to 53.06\% in a poorly tested project (poorly tested in regard to the mutation score).
Furthermore, based on the column \texttt{\#MA} we see that these improvements are made mostly through the addition of just a few test methods.

\subsection{Lessons learnt}

Next, we investigated the classes where we did not see any improvement.
In this section we share what we learnt from this investigating:

\paragraph{Amplification of pickles is challenging.}
Python has the ability to (de-)serialize the structure and states of objects, whereby the object is converted into a bitstream or vice-versa.
This makes it possible to read and write objects to files or send them across a network.
The process is called Pickling and makes use of the \texttt{pickle} module.
It is very important when trying to pickle and unpickle an object that it is imported the same way.

Since AmPyfier works based on mutation testing and the unit under test may be mutated, pickling can cause exceptions during the execution of the test method, either during observation or selection.
That is the main reason we mostly do not see any improvement in test methods that specifically test the pickling of an object.

We encountered this problem in two test classes, namely \texttt{DictTests} and \texttt{ChildDictTests} of the project Addict.
However pickling is only tested in one of the 63 methods, so increase in mutation score was still possible for both classes.

\paragraph{Amplification of file-based tests is not effective.}
A similar situation happens when the test works based on file inputs.
In such a test method, a file is opened and the object under test is initialized based on its content.
In these cases, current input amplification operators are not efficient, because they only consider the test source code and generate new inputs by mutating it.
They can not manipulate the content of the files to force the test to new states.

For example, in the test class \texttt{PdfReaderTestCases}, one of the test methods opens a PDF file and initializes a \texttt{PdfFileReader} object from the content of the file.
Then it uses the initialized object to assert some values.
Literal input amplification operators in this test have low effectiveness because they mutate the string of the file path.
Such mutations only generate an incorrect file path, which causes the raising of an exception, and does not permit AmPyfier to explore the search space of possible test inputs.

In total, three test classes had this problem, namely the above mentioned \texttt{PdfReaderTestCases} and \texttt{AddJsTestCase} of the PyPDF2 project and \texttt{TestMPQArchive} of MPyQ.
In only one of those three classes we were able to increase the mutation score.

\paragraph{Smarter assertion generation for complex objects is needed.}
In assertion amplification, AmPyfier needs to convert the observed object/value into an AST representing that object.
This process is fairly simple for asserting literals, or lists/dictionaries consisting of literals.
However, for complex objects this more challenging.
DSpot and \sa, record the values from getters and public attributes of each object.
If the value is also another object, it repeats the process for it up to a defined depth.
However, this technique easily clutters the test with a large amount of extra assertions, rendering the test unreadable.

For example, suppose generating assertions for a Matrix implementation.
The observer based on public getters observe  \texttt{matrix[1,1]} in depth 1, \texttt{matrix.row(1)[1]}, \texttt{matrix.column(1)[1]} and \texttt{matrix.transpose()[1,1]} in depth 2, and \textellipsis, while all these observations refer to a single element.
As a result, plenty of unnecessary assertions are generated.
Hence, it is important to make assertion amplification module more smarter.

Additionally, some developers define customized assertions, or helper methods that contain a set of assertions for specific types.
The test generator should be consistent with such a coding style, which it currently doesn't do.
A possible solution is using the profiling module to profile how an object is asserted in the existing test suite.
Using this mechanism, the assertion amplification module would learn from developers how to assert objects and would produce similar code.

\paragraph{Helper methods needs to be considered.}
Developers frequently use helper methods in their tests.
A helper method is another method defined in the test suite, but it is not a test method.
It can be defined for different reasons like grouping a set of assertions to reuse them in different test methods, or setting up an object and bringing it to a particular state.

If a helper method includes assertion statements, it is necessary to consider stripping them before input amplification.
In our experiments, for example the class \texttt{AddJsTestCase} uses such helper method.
As a result, the amplification of this class has not been successful, because the majority of input amplified tests were failing.
Helper methods that are defined for altering inputs also need to be considered in input amplification for increasing space exploration.

Helper methods were only used in the above mentioned \texttt{AddJsTestCase}.

The presence of a helper method is one of the main differences between a test generation and a test amplification tool.
In test generation, the tool starts to generate the tests from scratch and the structure of a test is defined by the tool.
However, in test amplification, the tool needs to adopt the style of the existing test suite and consider its elements.

\paragraph{Idiomatic Python Code}

Test amplification is a program synthesis task which generates new test methods based on existing ones.
The generated test methods needs to be merged into the code base to make the effect of improved tests permanent.
Therefore, it is important that the generated code is readable and adheres to the idiomatic coding conventions~\cite{knupp2013writing}.

In the current implementation of AmPyfier, the test method are changed via input amplification operators which make subtle changes to the original test code.
So, if the original test code follows Python coding idioms, the transformed code is likely to follow the same idioms.
Additionally, when generating new method calls and asserting values, AmPyfier respects the Python conventions in accessing private and protected members.
If we decide to extend Ampyfier with more complex input amplifiers, then the idiomatic nature of the code transformation should be taken into account.

Nevertheless, AmPyfier currently does not choose meaningful names for the generated test methods, as for instance done by Nijkamp et. al.~\cite{Nijkamp2021amplifiedTestNames}.
The tool also adds plenty of assertions, some of which are irrelevant, which has a negative effect on the readability of the generated code.
So, the produced test code is not immediately ready to be merged to the code base and it need a revision by a developer.

\paragraph{Continuous test amplification may help}

Lastly, as already briefly touched upon, the main downside to AmPyfier, and Test Amplification in general, is its complexity.
It is a time-, and resource-consuming task.
At worst, each test has to be executed the amount of mutants times the amount of amplified test generated times, to score them all.
Furthermore, multiple runs of the test are needed during assertion amplification to evade flaky tests, or to observe tests after an exception occurred.
This makes it difficult to introduce test amplification into realistic devlopment environments.

Therefore, we suggest employing AmPyfier in the CI pipeline to be triggered periodically or in each code push to amplify the recent changes.
In this case, the mutants would be generated only for the changed portions of code, and consequently the amplified tests should cover the changes.
This can significantly reduce the load of execution.
The amplified tests and other reports can be exported as artifacts, or they can be automatically sent as pull requests to be reviewed by the developers.

\section{Conclusion}
In this paper, we introduced AmPyfier, a test amplification tool for Python, a dynamically typed and interpreted language.
To built the tool we heavily relied on the design of DSpot, a test amplification tool for the statically typed language Java.
Yet, to overcome the shortcomings of a dynamically typed language, we incorporated type profiling as used by \sa for the dynamically typed language Smalltalk.
For Python specifically, we leveraged the fact that it is an interpreted language, where lots of information is available at runtime via reflective facilities such as the \texttt{sys.settrace()} function and the \texttt{inspect} module.
AmPyfier introduces a way to decrease the runtime of test amplification for large projects through the usage of multi-level metrics, where we amplify against code coverage before amplification on mutation score.

We evaluated AmPyfier on 7 open-source projects, and found that the tool could successfully strengthen 7 out of 10 test classes (70\%).
This is comparable to the results of DSPot (successful  in 65\% of the cases) and \sa (successful  in 60\% of the cases).
We collected qualitative evidence from the cases where Ampyfier failed to strengthen the test suite, deriving lessons learned and sketching areas for further improvement.
Despite these shortcoming, we conclude that test amplification is feasible for one of the most popular programming languages in use today.

\vspace{0.5cm}

\section*{Acknowledgments}
\ifthenelse{\boolean{anonymous}} {
\small{This work is sponsored by the ANONYMOUS of COUNTRY under a project entitled ``ANONYMOUS".}
} {
This work is supported by (a) the Fonds de la Recherche Scientifique-FNRS and the Fonds Wetenschappelijk Onderzoek - Vlaanderen (FWO) under EOS Project 30446992 SECO-ASSIST (b) Flanders Make vzw, the strategic research centre for the manufacturing industry.

}


\nocite{*}
\bibliography{bib}%

\end{document}